\documentclass{article}
\usepackage{amsmath}
\usepackage{amsfonts}
\usepackage{epsfig}
\usepackage{amssymb}
\usepackage{a4wide}
\usepackage[T1]{fontenc}
\usepackage{textcomp}
\usepackage{mathptmx}
\def\FF{\hbox to 8.33887pt{\rm I\hskip-1.8pt F}}
\def\NN{\hbox to 9.3111pt{\rm I\hskip-1.8pt N}}
\def\PP{\hbox to 8.61664pt{\rm I\hskip-1.8pt P}}
\def\QQ{\rlap {\raise 0.4ex \hbox{$\scriptscriptstyle |$}}
{\hskip -4.5pt Q}}
\def\RR{\hbox to 9.1722pt{\rm I\hskip-1.8pt R}}
\def\ZZ{\hbox to 8.2222pt{\rm Z\hskip-4pt \rm Z}}

\def\lbt{\left(}
\def\rbt{\right)}

\def\det{{\rm det}}

\newcommand{\resetequ}{\setcounter{equation}{0}}
  
\newcommand{\cA}{{\cal A}}            

\newcommand{\be}{\begin{equation}}
\newcommand{\ee}{\end{equation}}
\newcommand{\bqa}{\begin{eqnarray}}
\newcommand{\eqa}{\end{eqnarray}}
\newcommand{\ba}{\begin{array}}
\newcommand{\ea}{\end{array}}

\newcommand{\al}{\alpha}

\newcommand{\De}{\Delta}

\begin{document}
\title{Non-Commutative Renormalization}
\author{V. Rivasseau and F. Vignes-Tourneret\\
Laboratoire de Physique Th\'eorique\\
Universit\'e Paris-Sud XI\\
Orsay Cedex France}
\maketitle

\begin{abstract}
We review the recent approach of Grosse and Wulkenhaar to the 
perturbative renormalization of non commutative field theory and suggest a 
related constructive program. This paper is dedicated to J. Bros on his 65th birthday.
\end{abstract}

\section{Introduction}

Non-commutative field theory has attracted interest in recent years from several different point of views
\cite{DouNe}.
It is a concrete example of non-commutative geometry which, according to mathematicians
like A. Connes, should be in a broad sense the correct framework to quantize gravity.
It can be generated as an effective limit of string theory in certain cases. And it is related
to condensed matter problems such as two dimensional Fermi systems in the presence of strong magnetic transverse fields, i.e. the quantum Hall effect.
 
Quantizing gravity is considered the Graal of theoretical physics, and should 
lead to new insights on the ultimate nature of space, time, and the universe. In this direction 
a more and more insistent thread of the recent years is the possibility of some duality or mixing
between short and long distance physics. 
A priori this idea runs against the well
established tradition of determinism, the modern embodiment of which
is the standard philosophy of the renormalization group (RG). In this philosophy, iterated integration over
short range degrees of freedom of a bare "fundamental" microscopic action 
leads to an effective action for macroscopic long range variables. 
Infrared/ultraviolet duality or mixing clash with this philosophy but would fit well within the context of
increased links between particle physics
and astrophysics or cosmology. These links are certainly a major trend of physics in the last 
decade, which saw supernovas and cosmic background radiation studies boost
the concept of the universe and the big bang as the ultimate laboratory and experiment. 

A paradigmatic example for the possibility
of infra-red/ultraviolet mixing 
is the $R \to 1/R$ duality for strings compactified on circles of radius $R$.
If something of this kind occurs in the real universe, we might have to change the usual renormalization group picture in which microscopic observations solely determine fundamental laws.
We could view ourselves as observing the universe from some kind of a "middle scale", which could 
be called the "ultrared-infraviolet". Our observations of microscopic ultraviolet scales through e.g. accelerators and of macroscopic infrared scales through various kinds of "telescopes" might more and more reveal entangled aspects of physics. Both
should be considered inputs of some ultraviolet-infrared bare laws of physics and then 
combined into a modified RG analysis that would lead to effective laws of physics (and perhaps
of chemistry and biology as well?) for the ultrared-infraviolet middle scale. This may smell like the "anthropic" point of view. But what are the scales involved in this grand mixing?
The quantization scale for gravity is the Planck scale of about $10^{-33}$ meters. The observable present radius of the universe is about ten giga light-years or $10^{26}$ meters. Although
there may be nothing fundamental with this value, if we follow
the fashionable guess that new physics just lies "a factor of 10 around the corner of present observations", we obtain 
$10^{27}$ meters as a possible fundamental macroscopic length. A geometric ratio betwen these microscopic and macroscopic scales leads
to a fundamental "middle scale" of about a millimeter. Therefore we should perhaps call this point of view 
the "antropic" rather than anthropic principle... 

More seriously, from the mathematical physicist point of view, it seems a long way before a completely
rigorous string or $M$-like version of quantized gravity can be developed. In the mean time, it would be good to have simpler mathematical toy models of ultraviolet-infrared mixing in which we could begin to grasp the phenomenon. On this road simple Lagragian non-commutative field theories stand as the
first natural station.

Indeed the simplest way to generalize the algebra of ordinary commuting space time coordinates
is to add a non zero constant commutator between them. This means the relation
\be   [x ^{\mu}, x^{\nu}]  = i \theta^{\mu \nu}
\ee
should hold, with $\theta $ a constant antisymmetric tensor. As in symplectic geometry we could by a linear transformation
put $\theta$ under standard symplectic form. In this way space-time coordinates would occur as "symplectic pairs"
and we would have the simple notion of non-commutative $\RR^2$ or $\RR^4$ for which 
$\theta^{12} = \theta^{34} = -1$, $\theta^{21} = \theta^{43} = +1$ and all other $\theta^{\mu \nu}$'s
are zero\footnote{Of course we can also consider e.g. a non-commutative $\RR^3$ in which there
is just one symplectic pair and a remaining commutative coordinate but this is not very appealing.}.
Obviously this new tensor breaks Lorentz invariance. 

It also breaks locality. Indeed the ordinary product of functions is replaced by a 
non-commutative product called star product or Moyal product.
By the Campbell-Hausdorff formula we can compute the star product of two plane-waves:
\be  
e^{ik.x} \star e^{ik'.x}  =  e^{- \frac{i}{2}k_{\mu}\theta^{\mu \nu} k'_{\nu}} e^{i(k+k').x} \label{starplane}
\ee
from which one can deduce by linearity and Fourier analysis a kernel which explicitly displays the non-locality of the star multiplication:
\be  f \star g (z) = \int dx dy K(z ; y, x) f(x) g(y)\ ,
\ee
\be  K(z ; y, x) = \delta(z-x) \star  \delta(z-y) = {1\over (2\pi)^d}   \int dk dk' e^{ik(z-x)} \star e^{ik'(z-y)}
= {1\over (2\pi)^d  \det \theta}    e^{i(z-x)\theta^{-1} (z-y)}  .
\ee

\section{The UV-IR Problem for $\phi^4_4$}
\resetequ

The ordinary action for the simplest commutative renormalizable bosonic field theory,
the $\phi^4_4$ theory is:
\be  S = \int d^4 x \frac{1}{2}( \partial ^{\mu } \phi \partial_{\mu} \phi + m^2 \phi^2)  + {\lambda \over (4!)} \phi^4 \ ,
\ee
which is easily generalized to a naive non-commutative $\phi^4_4$ theory
by replacing ordinary products by the star product and integration by the non-commutative integration noted $\int_{nc}$ which is a kind of combination of ordinary integration plus a trace \cite{DouNe}:
\be
S = \int_{nc} \frac{1}{2}( \partial ^{\mu } \phi \star \partial_{\mu} \phi + m^2 \phi \star \phi)  + {\lambda \over (4!)} 
\phi \star \phi \star \phi \star \phi \ .
\label{naive} 
\ee
If we rewrite this action in momentum space and use the rule (\ref{starplane}) 
for the star product of plane waves, we get
\be
S = \frac{1}{2} \int  \frac{d^4p }{(2\pi)^4}
( p ^{2} + m^2) \phi (p) \phi (-p) + {\lambda \over (4!)} \int  \prod_{i=1}^4\frac{d^4p_i }{(2\pi)^4}  (2\pi)^4 
\delta(\sum_{i=1}^4 p_i)  \exp(-\frac{i}{2}   \sum_{i<j}   p_i^{\mu}  \theta_{\mu\nu} 
p_j^{\nu})  \prod_{i=1}^4 \phi (p_i) \; .
\ee
Remark that this formula is identical to the commutative case except for the oscillating vertex factor 
\be
\exp(-\frac{i}{2}   \sum_{i<j}   p_i^{\mu}  \theta_{\mu\nu} p_j^{\nu} ) \ . \label{oscill}
\ee 
The form of this factor shows that 
we cannot order fields arbitrarily at a vertex but that we have to respect cyclic ordering, and that the Feynman graphs of this theory are ribbon graphs, as in matrix models. In fact this {\it is}
a matrix model if we introduce the correct functional basis. Let us recall that experts 
concluded that the model is not renormalizable because of infrared/ultraviolet entanglement \cite{MiRaSe}-\cite{CheRoi}. 

Indeed planar graphs have amplitudes identical to ordinary $\phi^4$ graphs, since the oscillating factors
all cancel out in the planar case. Still this means the theory has an infinite number of divergent graphs
which require renormalization. But on the other hand for non-planar graphs the oscillating factors (\ref{oscill}) make the amplitudes finite but divergent when external momenta become small. This
leads in turn to infrared divergences of arbitrary strength when these non planar subgraphs are inserted into loops. No simple idea solved this "entanglement".

For instance for the planar tadpole graphs of Figure 1A, the amplitude is the standard diverging
one: 
\be   \frac{\lambda}{6} \int  \frac{d^4k}{(2\pi)^4}   \frac{1}{k^2 + m^2}\ ,
\ee
whether for the non planar tadpole of Figure 1B, it is
\be \frac{\lambda}{12} \int \frac{d^4k}{(2\pi)^4)} \frac{e^{ip^{\mu} k^{\nu} \theta_{\mu\nu} }}{k^2 + m^2} 
=  \frac{\lambda}{48\pi^2}  \sqrt{\frac{m^2}{\tilde p^2}}  K_1(\sqrt{m^2  \tilde p ^2})\ ,
\ee
where $K_1$ is a standard special function and $\tilde p_{\mu} = \theta_{\mu \nu} p^{\nu}$.
This amplitude behaves as $\tilde p ^{-2}$ for small $p$.
It does not seem possible to improve such a finite amplitude by an ultraviolet subtraction.
Furthermore if we insert $n$ of these non planar tadpoles in a closed loop as in Figure 2, we get
divergence of the integral at small $p$ of arbitrarily high power when $n$ gets large.

\begin{figure}
\centerline{\epsfig{figure=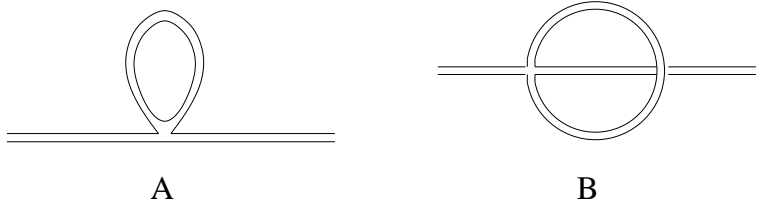,width=10cm}}
\caption{Planar and twisted tadpoles}
\centerline{}
\end{figure}

\begin{figure}
\medskip
\centerline{\epsfig{figure=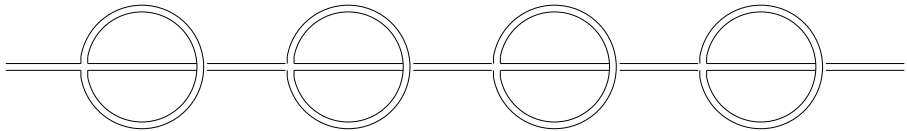,width=12cm}}
\caption{Chain of twisted tadpoles}
\end{figure}

\section{The Covariant Theory}
\resetequ

The solution to this problem was recently obtained by Grosse and Wulkenhaar. In a beautiful series of papers \cite{GrWu1}-\cite{GrWu2}-\cite{GrWu3}-\cite{GrWu4}, they claim that action (\ref{naive}) is indeed not renormalizable but simply 
in the sense of an anomalous theory, because it is not covariant under 
the Langmann-Szabo duality \cite{LaSz}. They found the correct Langmann-Szabo covariant
version of the theory by modifying the quadratic part of the action, and proved the BPHZ theorem, namely renormalizability of the theory at all orders in perturbation. Their proof still assumes some reasonable bounds on the propagator that they
checked numerically. We have now proved these bounds using a different type of cutoffs 
\cite{RiViWu}, so that the proof is complete even for "purists".

To understand why action (\ref{naive}) 
is not correct but anomalous we should introduce the Langmann-Szabo
duality:
\be   p_{\mu}    \to \tilde x _{\mu} = 2 (\theta^{-1})_{\mu \nu} x ^{\nu}\ ,
\ee
and define the Fourier transform of the field with a cyclic sign:
\be  \hat \phi (p) = \int d^4 x  e ^{(-1)^a  i p_{a,\mu} x_a^{\mu}} \phi(x_a)   \ , 
\ee
where the index $a=1,2,3,4$ follows a cyclic ordering at the vertex.

Then if we write this interaction in terms of ordinary products and kernels:
\bqa  S_{int} = \int_{nc}  \phi \star \phi\star \phi\star\phi   (x)
&=& \int  \prod_{a=1}^4 d^4x_a \phi(x_1) \phi (x_2) \phi(x_3) \phi   (x_4) V(x_1, x_2,  x_3, x_4)
\nonumber\\
&=& \int  \prod_{a=1}^4  \frac{d^4 p_a}{(2\pi )^4} 
\hat \phi(p_1)\hat \phi (p_2) \hat \phi(p_3) \hat \phi   (p_4) \hat V(p_1, p_2, p_3, p_4)
\eqa
we find 
\be V(x_1, x_2,  x_3, x_4) =\frac{1}{\pi ^4 \det \theta } \delta^4 (x_1 - x_2 + x_3 - x_4) \cos 
[ 2 \theta^{-1}_{\mu \nu} (x_{1\mu}x_{2\nu}  + x_{3\mu}x_{4\nu}]
\ee
\be \hat V(p_1, p_2, p_3, p_4) = (2\pi )^4 \delta(p_1 - p_2 - p_3 + p_4) \cos \frac{\theta^{\mu \nu}}{2}
[p_{1\mu}p_{2\nu}  + p_{3\mu}p_{4\nu}]
\ee
and this interaction has Langmann-Szabo duality.
But the free part of the action is not covariant under this duality; the local mass term,
like the interaction, is covariant
but the $p^2$ Laplacian is not because it has no $\tilde x^2$ counterpart. 
So Grosse and Wulkenhaar proposed as correct Langmann-Szabo covariant 
action for non commutative $\phi^4$ the generalized action with an harmonic potential dual to the kinetic
term:
\be  S_{free} = \int_{nc} \frac{1}{2} [ \partial_{\mu} \phi \star \partial^{\mu} \phi + \Omega^2 ( \tilde x_{\mu} \phi \star \tilde x^{\mu} \phi   + m^2 \phi \star \phi  ] \ .
\label{actcorr}
\ee
Then 
\be
S_{free} + S_{int} = \int_{nc} \frac{1}{2} [ \partial_{\mu} \phi \star \partial^{\mu} \phi + \Omega^2 ( \tilde x_{\mu} \phi \star \tilde x^{\mu} \phi   + m^2 \phi \star \phi  ] + {\lambda \over (4!)} 
\phi \star \phi \star \phi \star \phi 
\label{fullactcorr}
\ee
is covariant under duality $p_{\mu} \to \tilde x_{\mu}$:
\be  S(\phi , m, \lambda, \Omega) \to \Omega^2 S(\phi , \frac{m}{\Omega} , \frac{\lambda}{\Omega^2}, \frac{1}{\Omega}) \ .
\ee

The theory has one additional parameter, namely $\Omega$. It is completely invariant
at $\Omega =1$, where there is complete equivalence between space coordinates
and momenta.
It is this theory that is renormalizable at all orders of perturbation theory \cite{GrWu3}.

The best rigorous understanding of renormalization is neither in pure space or pure momentum
space, but in phase-space, the only frame in which it can be understood even in a constructive or non-perturbative sense.
But what is the equivalent of phase space here? An important observation is that in 
a non commutative space one cannot simultaneously localize all $x$ coordinates with arbitrary 
precision. There should be analogs of the Heisenberg uncertainty relations but now between coordinates. This may be grasped using Gaussian wave packets. The star multiplication of 
such Gaussians is again Gaussian, like in the commutative case. But in contrast with that
commutative case, if we try to focus more and more a Gaussian by star multiplying with itself,
there is a limiting scale at which this focusing process must stop.
More precisely we have for a two dimensional symplectic pair with commutator
$[x^1 , x^2]= \theta$
\be  e^{- \frac{x^2}{ a^2}}  \star  e^{- \frac{x^2}{ b^2}}  = c e^{-\frac{x^2}{d^2}}
\ee
with $c= [1+ \theta^2 /(ab)^2 ]$, $d= \sqrt{\frac{a^2 b^2 + \theta^2}{a^2 + b^2}}$,
so that $d<a$ if $a > \sqrt \theta$, $d>a$ if $a < \sqrt \theta$ 
and there is a fixed point at $a = \sqrt \theta$.

This means there is a canonical best focused Gaussian which is
\be f_0 (x_1, x_2) = 2 e^{- \frac{x_1^2 + x_2 ^2}{\theta}} .
\ee

This suggests to introduce creation and annihilation operators
\be  a = \frac{x_1 + i x_2}{\sqrt 2}   ; \ \bar a = \frac{x_1 - i x_2}{\sqrt 2}
\ee
and the base functions
\be   f_{mn} (x) = \frac{1}{\sqrt{n!m! \theta^{m+n}}}    \bar a ^{\star m} \star f_0 \star a ^{\star n}
\ee
in which the star product becomes an ordinary matrix product:
\be  f_{mn} \star f_{kl} = \delta_{nk} f_{ml} \ .
\ee

This is called the matrix base. Any function of $x$ or its Fourier transform in $p$ which belongs 
to Schwartz space, hence is smooth  with rapid decrease, can be analyzed as a series
in the $f_{mn}$ modes:
\be   a(x) = \sum_{m,n =0}^{\infty}  a_{mn}   f _{mn} (x) \ ,
\ee
\be   a \star b (x) = \sum_{m,n = 0}^{\infty}  (ab)_{mn} f_{mn} (x)
\ \ ; \ \    (ab)_{mn}  = \sum_{k=0}^{\infty} a_{mk} b_{kn}\ .
\ee
The condition of smoothness and rapid decay is equivalent to rapid decay of
the $a_{mn}$ coefficients:
\be
\forall k \ \ \sum_{m,n}   \bigl[  (2m+1)^{2k}   (2n+1)^{2k} \vert a_{mn} \vert ^2  \bigr]^{1/2} < \infty \ ,
\ee
and the non commutative integral becomes an ordinary trace:
\be
\int_{nc}  f_{mn} (x) = 2 \pi \theta \delta_{mn} \ .
\ee

In this theory, to cut  the large values of $m$ and $n$ for instance through 
a condition such as $m, n < \Lambda$
corresponds to impose a cutoff in both position and momentum space, so it is an infrared-ultraviolet
cutoff and the remaining modes $m, n < \Lambda$ could be called the ultrared-infraviolet modes below
$\Lambda$. As we said above the usual RG philosophy is now to integrate slice by slice the 
high $m,n$ modes to get the effective theory for the remaining ultrared-infraviolet modes. The renormalization condition instead of being formulated at zero or low momentum 
are now formulated at the fundamental mode $m=n=0$, which means the fundamental $f_0$ Gaussian
wave packet.

It is well known that  in practice there are different technical ways to implement this RG
philosophy, and in particular to create the different "momentum slices" for the "fluctuation fields" 
that are to be integrated out. The best way to do this when we have a positive quadratic
form as free action is to invert it and cut the slices directly on the propagator. 
This leads to factor the Gaussian measure into independent 
measures so that the slice fields are orthogonal with respect to this
decomposition. 

This is what Grosse and Wulkenhaar do in \cite{GrWu2}-\cite{GrWu3} 
For that they have first to compute the free action (\ref{actcorr}) in the matrix base,
which is not too difficult. One can perform first the computation for a single pair
of coordinates, so in $\RR^2$ since the two pairs are in fact independent. The result is
\be  S_{free} = 2 \pi \theta \sum_{m,n,k,l \in \NN}    \frac{1}{2} G_{m,n;k,l} \phi_{mn} \phi_{kl} 
\ee
\bqa  G_{m,n; k, l}  &=& m^2 +  \frac{2}{\theta}   (1 + \Omega^2) (n+m+1) \delta_{nk} \delta_{ml}
\nonumber\\
&-&  \frac{2}{\theta}   (1 - \Omega^2) (\sqrt{(n+1)(m+1)}\delta_{n+1, k} \delta_{m+1,l} + \sqrt{nm}
\delta_{n-1,k} \delta_{m-1,l} )\ .
\eqa
Then one has to invert this quadratic form to get
the propagator   of the theory in the matrix base. 
This is a beautiful non-trivial computation of Grosse and Wulkenhaar, which uses 
Meixner polynomials \cite{GrWu3}-\cite{KoSw},
which are some kind of hypergeometric functions. But we are really interested in the scaling properties
of this propagator. It is not necessary to know anything about Meixner
polynomials or special functions of any kind to understand or use the result,
which is based an integral representation of the propagator analog to Feynman-Schwinger 
parametric representation. The propagator $\Delta_{m,n; k ,l}$
is zero unless $m+l = n+k$, so we can express it in terms of three
multi-integers  $\Delta_{m,m+\alpha; l+\alpha ,l}$, and the result is
\begin{align}
\label{eq:propinit}
\Delta (m, m+\alpha; l + \alpha, l) = \frac{(1+\Omega)^2}{4\Omega} \int_0^1 dz\,  z^{\frac{\mu_0^2 \theta}{8 \Omega}}%
  \frac{\theta }{8\Omega(1 + B)^2} &\sum_{u=0}^{\min(m,l)}%
  {\cal A}\ \lbt\frac{\sqrt{z}}{1+B} \rbt^{2u+\alpha} \times\nonumber\\
  &\lbt \frac{B (1+\Omega)}{(1 + B)(1-\Omega)} \rbt^{m+l-2u}  
\end{align}
where ${\cal A}(m,l,\alpha,u)=\sqrt{\binom{m}{m-u}\binom{m+\alpha}{m-u}\binom{l}{l-u}\binom{l+\alpha}{l-u}}$ and
$B(z, \Omega)= \frac{(1-\Omega)^2(1-z)}{4\Omega}$. Indices such as $m$, $l$, $\alpha$ and $u$ have two components, one for each symplectic pair of ${\RR}^4$, hence run over $\NN^2$.
Sums or products apply to these two components.

Remark that every contribution in these sums and integrals is positive, so we should 
think to this representation as the non-commutative analog of the heat kernel or path 
representation of the propagator in the commutative case.

Grosse and Wulkenhaar then go on, slicing the propagator with sharp conditions
on the maximum of the four indices $m,n;k , l$ appearing in the propagator.
For instance the $i$-th slice of the RG in their framework would be
\be  \Delta_i (m,n;k,l) = \Delta (m,n;k,l)  \chi ( M^i \le \max (m,n,k,l) < M^{i+1} )
\ee
where $M$ is some fixed number, e.g.  $M=2$, and $\chi (X)$ is the characteristic function 
of the event $X$ considered.

Then they check that the theory with full action is renormalizable, using the
popular Polchinski's scheme which feeds inductive bounds directly into the RG
equations \cite{Po}. This is a good way of proving perturbative renormalizability
and even large order bounds. 

Let us however recall that until now there is no constructive
version of this scheme. Indeed it seems very difficult to iterate
bounds in the RG expansions of constructive field theory because many 
potential Wick contractions remain hidden in functional integrals. For example in
the simplest cases of Fermionic models, constructive field theory
simply amounts to expand a tree of Wick contractions connecting external sources
and to keep half of the contractions (those of the loops) inside a determinant 
\cite{DR}. This determinant is bounded in the end through some Gram's inequality.
But intermediate bounds \`a la Polchinski on earlier kernels 
at an intermediate stage of the expansion seem to inevitably destroy the
cancellations in the determinant that are responsible for constructive i.e. 
non-perturbative convergence of the scheme.

\section{Smooth slices}
\resetequ

We would like to introduce an improvement on \cite{GrWu2}-\cite{GrWu3} original scheme, by using
sharp cutoffs in Feynman parametric space rather than directly
on the indices $m$ and $n$. As is known e.g. in constructive theory, sharp cutoffs 
in direct or momentum space for ordinary field theory have bad decay properties
in the dual Fourier variable, and this leads to difficulties for
phase space analysis, whence cutoffs implemented through the parametric representation
are usually much more convenient. So we suggest
to pick a number $M>1$ and to define the $i$-th slice of the RG analysis through the sliced
propagator
\be
 \De_i(m, m+\al; l + \al, l ) = \int_{1-M^{-i+1} }^{1-M^{-i} } 
dz  z^{\frac{\mu_0^2 \theta}{8 \Omega}}  \frac{\theta}{8\Omega(1 + B)^2} 
\{ \frac{\sqrt{z}}{1 + B} \}^{m+l +\al}
\sum_{u=0}^{\min(m,l)} \cA 
\{ \frac{B(1+\Omega)}{\sqrt{z}(1-\Omega)} \}^{m+l -2u} 
\ee 

We can then  use bounds such as

\be  \cA(m,l,\al,u) \le \frac{\sqrt{m(\al +m)}^{m-u}\sqrt{l(\al +l)}^{l-u} }{(m-u)!(l-u)!} 
\le\frac{ (m+\al/2)^{m-u} (l+\al/2)^{l-u}}{(m-u)!(l-u)!}
\ee
in the ultraviolet region, which  corresponds to $z$ near 1. 

Let us consider a slice $M^{-i} \le 1-z \le M^{-i +1}$.
For $i$ large enough we have 
\bqa   \De_i(m, m+\al; l + \al, l ) &=& \int_{1-M^{-i+1} }^{1-M^{-i} } 
dz  z^{\frac{\mu_0^2 \theta}{8 \Omega}}  \frac{\theta}{8\Omega(1 + B)^2}  \{ \frac{\sqrt{z}}{1 + B} \}^{m+l +\al}
\sum_{u=0}^{\min(m,l)} \cA 
\{ \frac{B(1+\Omega)}{\sqrt{z}(1-\Omega)} \}^{m+l -2u} \nonumber \\
&\le & \int_{1-M^{-i+1} }^{1-M^{-i} }   dz  z^{\frac{\mu_0^2 \theta}{8 \Omega}}  
\frac{\theta}{8\Omega(1+ B)^2} \{ \frac{\sqrt{z}}{(1 + B)} \}^{m+l + \al }
\nonumber \\ && 
\sum_{u=0}^{\min(m,l)} \frac{X^{m-u}  Y^{l -u} }{(m-u)!(l-u)!}
\eqa
where
$X= \frac{B(1+\Omega)(m+\al/2)}{\sqrt{z}(1-\Omega)}$ and
$Y = \frac{B(1+\Omega) (l+\al/2)}{\sqrt{z}(1-\Omega)}$.
Since obviously
$\sum_{u=0}^{\min(m,l)}  \frac{X^{m-u}Y^{l-u}}{(m-u)!(l-u)!}  \le \exp (X+Y)$:

\bqa   \De_i(m, m+\al; l + \al, l ) 
&\le & \int_{1-M^{-i+1} }^{1-M^{-i} }   dz  z^{\frac{\mu_0^2 \theta}{8 \Omega}}  
\frac{\theta}{8\Omega(1 + B)^2} \{\frac{\sqrt{z}}{(1 + B)} \}^{m+l + \al } \nonumber \\
&&   \exp  \{ \frac{B(1+\Omega)(m+l+\al)}{\sqrt{z}(1-\Omega)} \}
\eqa

Expanding to first order in $1-z = \epsilon$, since $\sqrt{z}\simeq 1-\epsilon/2$ 
and $B = C\epsilon  $ with $C(\Omega) = (1-\Omega)^2/4\Omega$, 
one finds, since $C+(1/2) - C\frac{1+\Omega}{1-\Omega} = \Omega/2 $. 
\bqa   \De_i(m, m+\al; l + \al, l ) 
&\le & \frac{K\theta}{\Omega}  \int_{1-M^{-i+1} }^{1-M^{-i} }  dz 
\exp  \{ (-C - 1/2 + C \frac{1+\Omega}{1-\Omega}  )\epsilon (m+l+\al)  \} \nonumber\\
&\le &  \frac{K}{\Omega} M^{-i}  \prod_{j=1,2} 
e^{-M^{-i} \Omega (\al + m +l )/2  }
\eqa
for some constants $c$ and $K$.

In the same vein, we shall prove in \cite{RiViWu} bounds such as 
\be   \sum_{l} \max_{\alpha} \ \De_i(m, m+\al; l + \al, l ) \le  \frac{K}{\Omega} M^{-i}  \prod_{j=1,2} 
e^{-cM^{-i} m}
\ee
for a certain range of parameters $\Omega$ including a neighborhood of $\Omega =1$.
These bounds establish the {regular non-local} matrix model behavior
in the sense of \cite{GrWu1}. This should complete for purists the corresponding renormalization theorem at all orders in perturbation theory of \cite{GrWu3}.

\section{Towards Non-commutative Constructive Field Theory}
\resetequ

We would like to extend these results of perturbative field theory to 
the non-perturbative or constructive level. For this we have roughly speaking
to perform the sum over all Feynman diagrams. 
This is usually possible for Bosonic models which are both stable and
asymptotically safe, and for Fermionic models which are asymptotically safe. 
As remarked already, one can no longer directly use Polchinski inductive scheme,
but there are similar constructive renormalization group schemes in phase space 
(see \cite{Riv1}-\cite{LNP} 
for reviews on general methods and results on constructive field theory).

\subsection{Gross-Neveu} 

One of the easiest models of constructive field theory is the Gross-Neveu model
of Fermions with $N$ colors and a quartic vector-like interaction.
In two dimensions this model has the same power
counting than $\phi ^4_4$. It shows asymptotic
freedom in the ultraviolet and mass generation in the infrared, all these features
having been rigorously established at the constructive level (\cite{DR}-\cite{KMR} and references therein). 

There is an analogous model on commutative $\RR^2$ which is studied at large
$N$ in \cite{AkDeSe}. However if we rewrite the model in the matrix base, the regime
of large matrix indices is not the one studied in  \cite{AkDeSe}. The non-commutative Gross
Neveu model on $\RR^2$ ought to be the easiest candidate for rigorous non-perturbative
construction of a non commutative field theory. 
But we expect at least one difficulty
which is the following. In Fermionic vector models the easiest constructive analysis of \cite{DR}
can be summarized as follows. Instead of developing the full set of Wick contractions
that leads to the series of all Feynman graphs, one develops only "half of them",
namely those Wick contractions of a spanning tree in each diagram, and the other contractions are kept
in a Grassmann unexpanded integral, i.e. a determinant. Gram's bound is then applied on this determinant, which does not generate the dangerous factorial leading usually to divergence
of perturbation theory. This scenario works well in vector theories, such as for instance
the interacting Fermi liquid in two dimensions \cite{DR1}-\cite{FKT}, because Gram's bound
leads to a single sum over vector index at each vertex.

Here however we have matrix indices for each field or anti-field. After picking an explicit tree a naive 
Gram's bound on the remaining determinant may lead to two index sums at each vertex,
which is one more that what power counting can afford. In short, in an $N$ 
vector model  Pauli's principle stops perturbation theory roughly at order $N$ 
but in an $N\times N$ 
matrix model, Pauli's principle stops perturbation theory only at order $\simeq N^2$. 
This is somewhat reminiscent of the situation in three dimensional condensed matter models.
Hence the constructive analysis of non-commutative models may require
some heavier techniques, in fact the
"Bosonic techniques" of multi-scale cluster expansions, 
and Hadamard's bound instead of Gram's bound, as in \cite{DMR}.

\subsection{$\phi ^4_4$ at $\Omega =1$}

The ordinary $\phi ^4_4$ model, as well-known is disappointing from the constructive point of view
because for the repulsive sign of the coupling constant it lacks ultraviolet asymptotic freedom
and for the attractive sign it lacks stability! 

However in the non-commutative case there should be ultraviolet-infrared mixing so we expect the
asymptotic freedom in the infrared regime to come partly to the rescue of the ultraviolet regime. 
At $\Omega =1$
there should be perfect symmetry between the infrared and ultraviolet side. Therefore we expect
the $\beta$ function of $\phi ^4_4$ should vanish at $\Omega =1$.
This has been checked at least at one loop in  \cite{GrWu4}.

This means that at $\Omega =1$ the "running coupling constant" does not run at all, and the
bare coupling therefore equals the renormalized one. The model is not asymptotically free
but asymptotically safe. As a result there is no Landau
ghost problem, and no reason for which this theory should not exist in a non-perturbative sense,
probably being the Borel sum of its perturbative series. The same difficulties than for the 
Gross-Neveu model should be also expected here, namely the
constructive analysis should overcome the difficulties linked to a matrix model. However 
building non-perturbatively a $\phi ^4_4$ theory, even of a non-commutative type, is
a tantalizing perspective. 
After so many years it would partly realize one of the dreams of the founding fathers
of constructive theory, although this non-commutative model 
should obviously not be expected to fulfill Wightman's axioms, in particular
Lorentz invariance.


\begin{thebibliography}{99}

\bibitem{DouNe} M. R, Douglas and N. A. Nekrasov, Non commutative Field Theory,
arXiv:hep-th0106048 

\bibitem{MiRaSe} S. Minwalla, M. Van Raamsdonk and N. Seiberg, Non commutative
perturbative dynamics, JHEP 0002 (2000), 020.

\bibitem {CheRoi} I. Chepelev and R. Roiban, Renormalization of quantum field theories
on noncommutative $\RR^d$. I: Scalars. JHEP 0005 (2000) 037; Convergence Theorems
for non-commutative Feynman graphs and renormalization JHEP 0103 (2001) 001

\bibitem{GrWu1} H. Grosse and R. Wulkenhaar, 
Power Counting Theorems for non -local matrix models and renormalization, arXiv:hep-th/0305066 

\bibitem{GrWu2} H. Grosse and R. Wulkenhaar, Renormalization of $\phi^4$ theory
on noncommutative $\RR^2$ in the matrix base, arXiv:hep-th/0307017

\bibitem{GrWu3} H. Grosse and R. Wulkenhaar, Renormalization of $\phi^4$ theory
on noncommutative $\RR^4$ in the matrix base, arXiv:hep-th/0401128

\bibitem{LaSz} E. Langmann and R.J. Szabo, Duality in scalar field theory on noncommutative phase space, Phys Lett. B {\bf 553}, 168 (2002).

\bibitem{RiViWu} V. Rivasseau, F. Vignes-Tourneret and R. Wulkenhaar, to appear.

\bibitem{KoSw} R. Koekoek and R. F. Swarttouw,
The Askey-scheme of hypergeometric orthogonal 
polynomials and its $q$-analogue, arXiv:math.CA/9602214

\bibitem{Po} J. Polchinski, Renormalization and Effective Lagrangians,
Nucl. Phys B {\bf 231}, (1984), 269.

\bibitem{DR} M. Disertori, V. Rivasseau, Continuous Constructive Fermionic Renormalization,
Ann. Henri Poincar\'e {\bf 1}, (2000), 1.

\bibitem{Riv1} V. Rivasseau, From perturbative to constructive renormalization, 
Princeton University Press (1991).

\bibitem{LNP} Constructive Physics, Proceedings of the International 
Workshop at Ecole Polytechnique, Palaiseau, July 1994, 
ed. by V. Rivasseau, Lecture Notes in Physics 446, Springer Verlag (1995).

\bibitem{AkDeSe} E. T. Akhmedov, P. de Boer and G. W. Semenoff,
Non Commutative Gross-Neveu Model at large $N$, hep-th/0103199

\bibitem{KMR} C. Kopper, J. Magnen and V. Rivasseau,
Mass Generation in the Large N Gross-Neveu Model,   
Commun. Math. Phys. {\bf 169} 121 (1995).

\bibitem{DR1} M. Disertori and V. Rivasseau, Interacting Fermi liquid 
in two dimensions at finite temperature, Part I and II, Commun. 
Math. Phys. {\bf 215}, 251 and 291 (2000).

\bibitem{FKT} Joel Feldman, H. Kn\"orrer and E. Trubowitz, A two dimensional Fermi Liquid, Commun. Math.Phys. {\bf 247}, 1-319, 2004 and Reviews in Math. Physics, {\bf 15}, 9,  949-1169, (2003).

\bibitem{DMR} M. Disertori, J. Magnen and V. Rivasseau, Interacting Fermi liquid
in three dimensions at finite temperature, part I: Convergent Contributions,
Ann. Henri Poincar\'e, {\bf 2} 733-806 (2001).

\bibitem{GrWu4} H. Grosse and R. Wulkenhaar, The $\beta$-function in duality-covariant
noncommutative $\phi^4$-theory, arXiv:hep-th/ 0402093

\end{thebibliography}
\end{document}